\documentclass[12pt]{article}
\usepackage{graphics,amsfonts}
\usepackage{epsfig}
\usepackage{latexsym}
\newcommand{\beq}{\begin{equation}}
\newcommand{\eeq}{\end{equation}}
\newcommand{\beqa}{\begin{eqnarray}}
\newcommand{\eeqa}{\end{eqnarray}}
\newcommand{\beqar}{\begin{eqnarray*}}
\newcommand{\eeqar}{\end{eqnarray*}}

\newcommand{\eg}{{\it e.g.,}\ }
\newcommand{\ie}{{\it i.e.,}\ }
\newcommand{\labell}[1]{\label{#1}} %{\label{#1}\qquad_{#1}} %
\newcommand{\reef}[1]{(\ref{#1})}

\def\IR{{\hbox{{\rm I}\kern-.2em\hbox{\rm R}}}}
\def\IZ{{\hbox{{\rm Z}\kern-.2em\hbox{\rm Z}}}}
\begin{document}

\thispagestyle{empty}
\rightline{\small hep-th/0008226 \hfill McGill/00-23; DTP/00/73}
\vspace*{2cm}

\begin{center}
{ \large \bf  Fractional Branes and the Entropy of 4D Black holes}
%\\[.25em]
%{ \LARGE and the Entropy of 4D Black holes}
\vspace*{2cm}

Neil R. Constable$^{a,}$\footnote{\tt constabl@hep.physics.mcgill.ca},
Clifford V. Johnson$^{b,}$\footnote{\tt c.v.johnson@durham.ac.uk} and
 Robert C. Myers$^{a,}$\footnote{\tt rcm@hep.physics.mcgill.ca}\\
\vspace*{1cm}
$^a${\it Department of Physics, McGill University}\\
{\it Montr\'eal, QC, H3A 2T8, Canada}\\
\vspace*{0.2cm}
$^b${\it Department of Mathematical Sciences, Durham University,}\\
{\it Durham, DH1 3LE, United Kingdom}\\
\vspace{2cm} 
\end{center}
\begin{abstract}
  We reconsider the four dimensional extremal black hole constructed
  in type IIB string theory as the bound state of D1--branes,
  D5--branes, momentum, and Kaluza--Klein monopoles. Specifically, we
  examine the case of an arbitrary number of monopoles. Consequently,
  the weak coupling calculation of the microscopic entropy requires a
  study of the D1--D5 system on an ALE space. We find that the
  complete expression for the Bekenstein--Hawking entropy is obtained
  by taking into account the massless open strings stretched between
  the fractional D--branes which arise in the orbifold limit of the
  ALE space. The black hole sector therefore arises as a mixed
  Higgs--Coulomb branch  of an effective 1+1 dimensional gauge
  theory.
\end{abstract}
\vfill \setcounter{page}{0} \setcounter{footnote}{0}
\newpage

\section{Introduction}

In the seminal work by Strominger and Vafa~\cite{stromv}, the then
newly discovered D--brane technology~\cite{joe} was used to give a
description of the Bekenstein--Hawking
entropy\cite{bekhawk} in terms of microscopic degrees of freedom associated
with open strings living on D--branes. This work considered
a class of extremal charged black holes in five dimensions.
Subsequent work by many authors extended this to include near
extremal black holes~\cite{cm,horstr}, and rotating black
holes\cite{jake1,jake2} in five dimensions as well as several black
holes in four dimensions~\cite{4d,sm,hlm}---see
refs.~\cite{juan,amanda,youm,bala} for details and extensive
references.

In the case of the five dimensional black holes of ref.~\cite{stromv},
there are three charges parameterizing the black hole, two from the
R--R sector and one from the NS--NS sector. These charges were
arranged in a very particular way in order to produce a black hole
with non--vanishing horizon area. It was noticed in
ref.~\cite{kallosh} that all solutions related to the solution of
ref.\cite{stromv} by the action of $E_{(6)6}$, the U--duality group in
five dimensions, also have a finite horizon area. This follows from the
fact that the horizon area and hence the entropy are related to the
cubic invariant of $E_{(6)6}$.  

When the type II theories are toroidally compactified to four dimensions,
the U--duality group is extended to $E_{(7)7}$ \cite{crm,hullt}.
One then finds \cite{kol} that the horizon area of four dimensional black
holes is naturally described by the quartic invariant of $E_{(7)7}$,
and hence a finite area requires the presence of at least {\it four}
charges. One interesting aspect of the enlarged U--duality group is that
such black holes can be constructed entirely from D--branes
\cite{fin,fm}. However, the first successful microscopic entropy
calculations were made in the context of systems carrying two NS
charges and two R--R charges \cite{4d,sm}. For example,
in ref.~\cite{4d}, the entropy of a four dimensional extremal
Reissner--Nordstrom black hole\footnote{This solution to four dimensional
Einstein--Maxwell gravity has been embedded into string theory 
in many different ways --- see, for example, 
refs.~\cite{kall,duff1,duff2,khuri,cvetic1,cvetic2,lu}.}
was computed with a
construction that was similar to that of the original five
dimensional work with the crucial addition of a fourth charge
corresponding to a Kaluza--Klein monopole. The calculations of ref.~\cite{4d}
considered the case when the charge of the monopole $Q_m$ was unity,
and so the symmetry between the four charges in $E_{(7)7}$ 
invariant was not manifest. In other approaches~\cite{sm,bala} related to
that presented in ref.~\cite{4d} by U-duality, which do not rely on
explicitly introducing a KK
monopole, the extra charge is easy to consider for values greater than 1. 
However, since those early days, the technology for studying
D--branes in non--trivial backgrounds, such as that created by a
Kaluza--Klein monopole, has improved considerably. It is now quite
easy to incorporate the case of a Kaluza--Klein monopole of charge
$Q_m > 1$, and we shall do that in this short note thus obtaining the full
expression for the entropy using the original set-up of ref.~\cite{4d}. 

To extend the results of ref.~\cite{4d} to the case $Q_m >1$ the main
property we need is that the core of the KK monopole is a Euclidean
Taub--NUT space \cite{gross}. Furthermore, the generalizations of
Taub--NUT spaces to include more than a single unit of
``NUT--charge'', are the multi--Taub--NUT metrics of
ref.~\cite{multihawk}. We shall need only consider the case where the
NUT centres are all coincident, in which case, close to the core of
the solution the space is simply the Gibbons--Hawking multi--centre
metric~\cite{GH} with coincident centres. The point in moduli space we
need will be the simple case where the space is nothing but the
$R^4/\mathbb{Z}_{Q_m}$ orbifold locally, which is a simple example of
an ``ALE space''.

The Bekenstein--Hawking entropy of the resulting four dimensional
black holes made by including the $Q_m>1$ monopoles will arise by
considering a modification of the D1--D5 bound state counting argument
to include the effects of being on the orbifold space time. The result
will turn out to be given in a simple manner by the product of four
charges, as one would expect if this is to be related to the quartic
invariant of $E_{(7)7}$. 

The point is simply to consider the D1--D5 bound state on an ALE
space.  The general theory of D--branes living on the full
``A--series'' spaces was first worked out in ref.~\cite{doug} and this
was extended to the full ADE family in ref.~\cite{ale}. (See also
refs.\cite{eric1, eric2, atish}.) In particular, it was found that
at an ALE singularity an individual D--brane can ``fractionate'' into
multiple D--branes (each carrying a fraction of the D--brane
charge)\cite{diacon} and can move independently of each other when
located at the singularity. It is by taking careful account of these
fractional D--branes which allows us to obtain the entropy as a
product of four charges, since the counting argument will inherit a
multiplicity coming from the $Q_m$--ness of the ALE space.

This paper is organized as follows. Since it has been a while that
these constructions have been discussed in the literature, in section
two we thoroughly review the construction of four dimensional charged
extremal black holes in supergravity and generalize these to cases
where $Q_m>1$. We observe that the Bekenstein Hawking entropy to be
given by the product of four charges. In section three we present the
D--brane construction of this black hole highlighting the necessary
details from the theory of D--branes on ALE spaces to show how this
leads to the result that the entropy is indeed the product of four
charges expected of an $E_{(7)7}$ invariant. We conclude with some
brief comments and discussion.

\section{Four Dimensional Charged Black holes}

It is well known that extremal black holes in four dimensions can be
constructed from supersymmetric bound states of ten (or eleven)
dimensional supergravity solutions, which are then dimensionally
reduced to $d=4$ ---for many examples and references
see~\cite{juan,amanda,youm,bala}. In type IIB supergravity, whose
bosonic fields are the metric $g_{\mu\nu}$, the antisymmetric
Kalb-Ramond field $B_{(2)}$ the dilaton $\Phi$ and the RR potentials
$C_{(0)},\,C_{(2)},\, C_{(4)}$ with the action\footnote{Here
$2\kappa^2=16\pi G_{10}=(2\pi)^7g_s^2l_s^8 $. This action should
also be supplemented with a self--dual five form field strength for
the RR potential $C_{(4)}$, but the latter will not play a role in the
following.}, 
\beq S=\frac{1}{2\kappa_{10}^2}\int
d^{10}x\sqrt{-g}\left(e^{-2\phi}\left[R+
    4\left(\partial\phi\right)^2-\frac{1}{12}\left(\partial
      B_{(2)}\right)^2\right] -\frac{1}{2}\left(\partial
    C_{(0)}\right)^2- \frac{1}{12}\left(\partial C_{(2)}-\partial
    B_{(2)}C_{(0)}\right)^2\right) \labell{action} 
\end{equation} 
this was first
considered in~\cite{4d}. There, the authors considered a bound state
of $Q_1$ D1--branes in the (09) plane , $Q_5$ D5--branes filling the
(056789) plane, momentum along the $x^9$ intersection of the D1--D5
world volume, and a Kaluza--Klein monopole with structure in the
(01234) directions.  The reduction to five dimensions is accomplished
by wrapping the (5678) directions on a $T^4$ with volume $V$ and the
$x^9$ direction on a circle of radius $R_9$. The final reduction to
four dimensions is performed on the $x^4$ circle which has length
$L_m$. The latter circle is non--trivially fibred over the $S^2$'s at
constant radius in the
$(x^1,x^2,x^3)$ directions in such a way as to construct a
Kaluza--Klein (KK) monopole.

Let us now review some of the features of the KK-monopole.
This monopole is a purely gravitational object and so appears
as a low energy solution in any string theory (including type IIB,
which is of interest here). The metric can be written as~\cite{gross}:
\begin{equation}
ds^2=-dt^2+\sum_{i=1}^5 dx_i^2+H_m\left(dr^2+r^2d\theta^2+
  r^2\sin^2\theta
  d\phi^2\right)+H_m^{-1}\left(dx_4+L_m{\bf A}\cdot{\bf dy}\right)^2
\labell{mono} 
\end{equation} 
where $H_m$ is a harmonic function given by, 
\begin{equation}
H_m=1+\sum_{j=1}^{Q_m}\frac{L_m}{|{\bf x}-{\bf x_j}|} \labell{harm}
\end{equation} 
here ${\bf A}$ is the electromagnetic vector potential in the 
$(t,r,\theta,\phi)$ subspace defined by,
\beq
\partial_{i}H_m\equiv -{\bf \nabla}\times {\bf A}
\labell{vec}
\eeq
In these expressions $Q_m$ is the number of 
monopoles whose
locations are given by the ${\bf x_j}$. 

The KK monopole considered in ref.~\cite{4d} had $Q_m=1$ for which the
$(r,\theta,\phi,x_4)$ subspace is exactly the Euclidean Taub--NUT
instanton solution, while for arbitrary $Q_m$ it is the
multi--Taub--NUT~\cite{TN,multihawk}.  For $Q_m=1$, in order to avoid
singularities at $r=0,\,\theta=0,\pi$ the coordinate $x_4$ must be
identified with periodicity $x_4\sim x_4+4\pi L_m$.  Consequently the
$(\theta,\phi,x_4)$ subspace is topologically an $S^3$. With the
above periodicity and generic choices of the parameters ${\bf x_j}$,
the solution \reef{mono} is everywhere nonsingular and 
corresponds to $Q_m$ separated,  charge one monopoles.
As we are interested here in constructing single centre black holes
carrying multiple monopole charges we will consider the special point
in the monopole moduli space with ${\bf x_j}=0, \forall j$ in which
case the harmonic function reduces to,
\begin{equation} 
H_m=1+\frac{Q_mL_m}{r}\ .
\end{equation} 

In the case of coincident monopoles with $Q_m>1$,
the $(r,\theta,\phi,x_4)$ subspace has a singularity at
the core, which will be of considerable use to us.\footnote{For
coincident centres, it is possible to adjust the periodicity of
$x_4$ to remove the ``NUT'' singularity, but it does not makes sense
to adjust this for each value of $Q_m$. Instead, we shall see that
there is a natural role for this singularity, as it endows the
D--branes with a multiplicity.}  The neighbourhood of the point
$r=0$ in fact looks locally like $R^4/\mathbb{Z}_{Q_m}$, a collapsed
$A_{Q_m-1}$--series ALE space. For arbitrary $r$, the $S^3$ we saw
previously is now in fact a $S^3/\mathbb{Z}_{Q_m}$. This is, like $S^3$, a
fibration of the $x^4$ circle over the $S^2$, but now there is an
action of $\mathbb{Z}_{Q_m}$ on the circle. When we construct the black hole
by including other harmonic functions, the singularity at $r=0$ is
resolved into an event horizon with finite area. The part of it in
four dimensions is of course the smooth $S^2$ over which the $x^4$ is
fibred.

In order to construct the black hole we apply the harmonic function
rule \cite{tsyt} to obtain the ten--dimensional string frame metric
representing the required  bound state:
\begin{eqnarray}
ds_{10}^2&=&H_{1}^{-1/2}H_{5}^{-1/2}\left(-dt^2+dx_{9}^{2}+H_p(dt-dx_9)^2\right)
+H_1^{1/2}H_5^{-1/2}\left(dx_5^2+dx_6^2+dx_7^2+dx_8^2\right) 
\nonumber \\
&+&H_1^{1/2}H_5^{1/2}H_m\left(dr^2+r^2d\theta^2+r^2\sin^2\theta d\phi^2\right)
+H_1^{1/2}H_5^{1/2}H_m^{-1}\left(dx_4+L_m\cos\theta d\phi \right)^2
\labell{tend}
\end{eqnarray}
while the dilaton $\Phi$ and the R-R three form field strength are,
\begin{equation}
e^{2\Phi}=\frac{H_1}{H_5}; \,\,\,\,\,\,\,\,\,\,\,
F^{(3)}=-d\left(\frac{1}{H_1}\right)\wedge dt\wedge dx_9+L_5\,
\sin\theta\,d\theta\wedge d\phi\wedge dx_4 %d\Omega_3
\labell{backgrd}
\end{equation}
In these expressions there are four harmonic functions which represent
the contributions of each of the constituents of the bound state.
Those of the D1--brane and D5--brane are given, (in the conventions of
ref.~\cite{juan}) respectively by,
\begin{equation}
H_1=1+\frac{8\pi^4g_s
\alpha^{\prime 3}Q_1}{VL_m}\frac{1}{r} \ ;\qquad
H_5= 1+\frac{g_s\alpha^{\prime}Q_5}{2L_m}\frac{1}{r}
\labell{har1}
\end{equation}
while those of the momentum wave and KK monopole are, respectively:
\begin{equation}
H_p=\frac{8\pi^4g_s^2\alpha^{\prime 4}N}{VR_9^2L_m}\frac{1}{r}\
;\qquad 
H_{m}=1+\frac{L_mQ_m}{r}\ .
\labell{har2}
\end{equation}
Here $V$ is the volume of the four torus making up the $(5,6,7,8)$
directions. Above, $L_5$ is the coefficient appearing 
in front of $1/r$ in the D5--brane
harmonic function. 

It is clear from the form of these harmonic functions that $r=0$ is
now a surface of finite area, containing the $S^2$ horizon of our four
dimensional black hole. Shifting to the ten dimensional Einstein frame
using $g_{\mu\nu}^{E}=e^{-\Phi /2}g_{\mu\nu}^{s}$, one can calculate
the area\footnote{Here we mean the area measured in ten dimensional
  units. } of the surface $r=0$ to be,
\begin{equation}
A=8\pi G_{10}\sqrt{Q_mQ_1Q_5N} \ ,
\labell{area}
\end{equation}
and thus the Bekenstein--Hawking 
entropy is \cite{bekhawk},
\begin{equation}
S=\frac{A}{4G_N^{10}}=2\pi\sqrt{Q_mQ_1Q_5N} \ .
\labell{entropy}
\end{equation}
which, as expected, is the product of the four charges characterising
the black hole obtained upon reduction to four dimensions.  Clearly
this reduces to the results of ref.~\cite{4d} for $Q_m=1$.

As a final comment on this solution, one might consider taking the near horizon
limit in order to examine the AdS/CFT duality~\cite{malda,review}. The
natural limit takes $\alpha^\prime\rightarrow0$ holding fixed $R_9$, $L_m$,
$u^2=4Q_mL_m\,r/\alpha^{\prime\,2}$ and
$\tilde{x}_i=x_i/2\pi\sqrt{\alpha^\prime}$ for $i=5,6,7,8$. The resulting
metric can be written in the following form,
\beqa
ds^2/\alpha^\prime&=&
\frac{u^2}{R^2}\left(-dt^2+dx_9^2+{u_p^2\over u^2}\left(dt-dx_9\right)^2\right) 
+\frac{R^2}{u^2}du^2
+\sqrt{\frac{Q_1}{v Q_5}}\left(d\tilde{x}_5^2+d\tilde{x}_6^2+d\tilde{x}_7^2+
d\tilde{x}_8^2\right)
\nonumber \\
&&\qquad\quad+{R^2\over4}\left[d\theta^2+\sin^2\theta\,d\phi^2+
\left(d\psi+1/Q_m\,\cos\theta\,d\phi\right)^2\right]
\labell{near}
\eeqa
where $R^2=g_s^2Q_mQ_1Q_5/vL_m$, $v=V/(2\pi)^4\alpha^{\prime\,2}$,
$u_p^2=g_s^2N/vL_mR_9^2$
and $\psi\equiv x_4/Q_mL_m$ with periodicity $\psi\sim\psi+4\pi/Q_m$.
The $(t,x_9,u)$ subspace
is simply $AdS_3$ in non-standard coordinates~\cite{juanandy} while the 
$(\theta, \phi,\psi)$ subspace is the coset $S^3/\mathbb{Z}_{Q_m}$.
Since this geometry is $AdS_3\times S^3/\mathbb{Z}_{Q_m}\times T^4$, it is 
dual to a two dimensional orbifolded conformal field theory\cite{ks},
which has been discussed in refs.~\cite{bern,suga}. In related work,
the quantization of strings in the Neveu--Schwarz background with this
geometry was discussed in ref.~\cite{leigh1}. In passing we note
that by slightly modifying the scaling limit (ie, scaling $L_m$ with $r$)
the KK-monopole function is retained in the near horizon limit. The
new metric becomes
\beqa
ds^2/\alpha^\prime&=&
\frac{u^2}{R^2}\left(-dt^2+dx_9^2+{u_p^2\over u^2}\left(dt-dx_9\right)^2\right) 
+\frac{R^2}{u^2}f_m(u)du^2
+\sqrt{\frac{Q_1}{v Q_5}}\left(d\tilde{x}_5^2+d\tilde{x}_6^2+d\tilde{x}_7^2+
d\tilde{x}_8^2\right)
\nonumber \\
&&\qquad\quad+{R^2\over4}\left[f_m(u)\left(d\theta^2+\sin^2\theta\,d\phi^2
\right)+f_m^{-1}(u)\left(d\psi+1/Q_m\,\cos\theta\,d\phi\right)^2\right]
\labell{nearb}
\eeqa
where $f_m(u)=1+u^2/Q_mL_m$. This new solution corresponds to perturbing
the dual CFT by an irrelevant operator of weight $\Delta=4$. 
For small $u$, we recover the $AdS_3$ geometry in eqn.~\reef{near}, that is
we recover the same CFT physics in the IR. However, the UV is strongly
modified as can be seen in eqn.~\reef{nearb} by the disruption of the
$AdS_3$ as $u\rightarrow\infty$. In this limit, the $\psi$-fibre in the
$S^3/\mathbb{Z}_{Q_m}$ part of the geometry shrinks to zero size, while
$S^2$ base expands to combine with the $du^2$ part of the line element
to yield the metric on $R^3$. It would be interesting to investigate
the physics of this perturbation further.

What we do next is to determine just how we can reproduce the above black
hole entropy \reef{entropy} from a microscopic description of the same system
at weak coupling in terms of a bound state of D--branes.

\section{Fractional Branes and Microscopic Entropy}

\subsection{The Roles of Higgs and Coulomb}

Specifically, we take a D--brane configuration consisting of $Q_1$
D1--branes extended in the (09) plane, bound inside the world--volume of
$Q_5$ D5--branes filling the (056789) directions where the (5678)
directions wrap a  $T^4$.  In ref.~\cite{4d} this configuration was
taken to be bound to a Kaluza--Klein monopole whose core lies in
the (01234) directions.  Further, the $x^9$ direction is
 an $S^1$ with radius $R_9$ and the D--brane configuration
is taken to have momentum $P^9=N/R_9$ in this direction. 

As we observed in the previous section, the neighbourhood of $r=0$ is
simply an A$_{Q_m-1}$--series ALE space,
\ie an $R^4/\mathbb{Z}_{Q_m}$ singularity appears at the centre of
the (1234) directions. The problem of calculating the
entropy of this configuration is thus the problem of counting the BPS
excitations of the D1/D5 system in the background of this ALE space.

As pioneered in ref.\cite{stromv}, the important features of the
bound state degeneracy problem are captured in the study of the
effective 1+1 dimensional gauge theory on the world volume of the brane
system. From the point of view of the D5--brane gauge theory, the
D1--branes are bound states in the ``Higgs branch'', in which the
D1--branes are ordinary instantons in side the D5--branes. This branch
is parameterised by the vacuum expectation values (vev's) of 1--5 open
strings, which give 4$Q_1Q_5$ bosonic and fermionic states, simply the
dimension of instanton moduli space.  The ``Coulomb branch'' of that
gauge theory is the situation where the D1--branes become point like
instantons and then leave the D5--branes\cite{edsmall}, ceasing to be
bound states. This branch is parameterised by the vev's of 1--1 and
5--5 strings, which ultimately separate the individual D--branes
from each other.  This takes us away from the black hole, the state of
most degeneracy.

The presence of the singular ALE space at the core of the monopole,
transverse to all of the branes introduces a new feature to the
problem. There will be, as we shall
see \cite{eric1,eric2,atish,doug,ale}, additional sectors in the
gauge theory when the branes are on the ALE space. These sectors also
have a Higgs and a Coulomb branch, this time both parameterised by
1--1 or 5--5 strings.  This time the branches represent the opposite
situation: When the D--branes (D1 or D5) are {\it on} the ALE singularity
($r=0$), and moving around ``inside it'', we are on the Coulomb
branch. There, all of the $Q_m$ monopole centres are at the origin: ${\bf
  x}_i=0$. The Higgs branch \cite{joetensor} is when the branes move
{\it off} the ALE space. There are also $Q_m$ parameters of the branch
corresponding to pulling apart the centres, making ${\bf x}_i\neq 0$.
So it is the Higgs branch which takes us away from the black hole,
this time.

Rather interestingly, then we see that we have a gauge theory with
distinct sectors (which we shall derive shortly) coming from with fact
that we have two sorts of brane, and also that they are both
transverse to the ALE space. The part of moduli space corresponding to
the black hole at strong coupling is a {\it mixed Higgs--Coulomb
  branch} of the gauge theory.\footnote{See ref.~\cite{cvj} for an
  extensive review of these Higgs and Coulomb branches of D--brane
  gauge theories, and their relevance to geometry, space time or
  otherwise.}

\subsection{The 1+1 Dimensional Gauge Theory}
\begin{figure}[ht!]
\begin{center}
\scalebox{0.7}[0.7]{\includegraphics{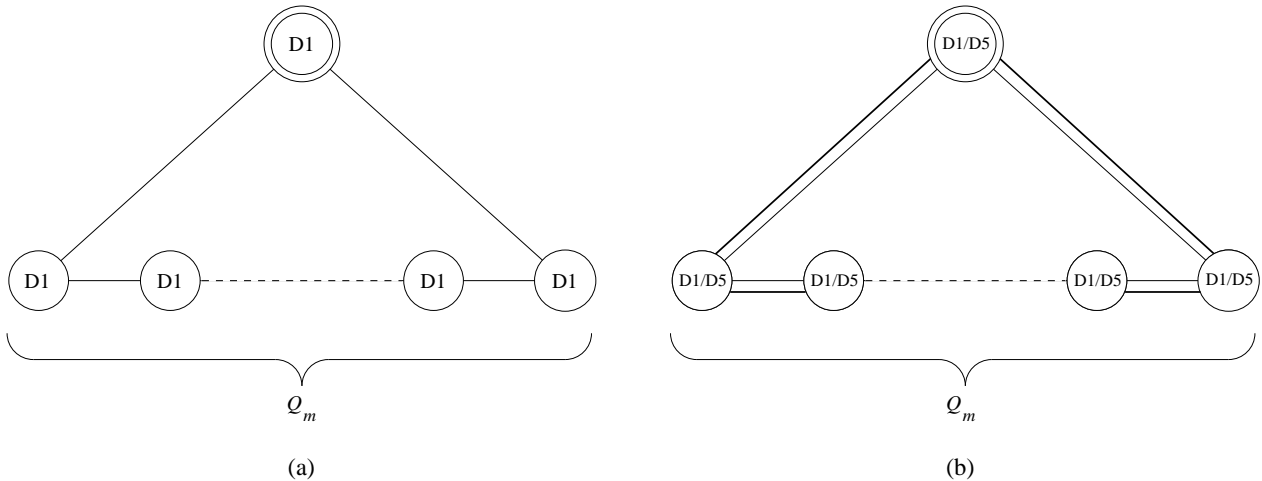}}
\caption{The extended Dynkin diagrams encoding the arrangement
  of open strings when (a) $Q_1$ D--strings and (b) $Q_1$ D--strings and $Q_5$
  D5--branes are placed at an $A_{Q_m}$ singularity. The heavy lines
  are $(5,5)$ strings and the lighter lines are $(1,1)$ strings. Each
  circle represents one factor of (a) $U(Q_1)$ or (b) $U(Q_1)\times U(Q_5)$.
  The double circle corresponds to the trivial representation.}
\label{fig:quiv}
\end{center}
\end{figure}

Let us review some aspects of the now standard construction of the
gauge theories on branes at orbifold
points\cite{doug,ale,eric1,eric2,atish,joetensor}. (See
ref.~\cite{cvj} for a more comprehensive review.) When a single
D--brane is brought to the orbifold $\mathbb{Z}_{Q_m}$ fixed point, one
works first on the covering space, introducing $Q_m$ images.  The
starting gauge group is therefore $U(Q_m)$, but after imposing the
orbifold projection, it is reduced to $U(1)^{Q_m}$. So the 1--1 open
strings living on the world--volume form a $U(1)^{Q_m}$ gauge field
and a hyper-multiplet in the adjoint of this gauge group, and
also a second family of massless hyper-multiplets transforming in the
``bi--fundamentals'' of pairs of $U(1)$ groups as in
figure~\ref{fig:quiv}. Each node in the diagram corresponds to a
$U(1)$ factor, and each link represents a hyper-multiplet with the
appropriate fundamental representation charges (1,--1). At the
ALE singularity, each
of the $Q_m$ nodes can be associated with a ``fractional'' D1--brane,
and the second set of hyper-multiplets are simply
the fundamental strings stretching between these
fractional D--strings as shown in the figure.
The Coulomb branch corresponds to giving the adjoint scalars an expectation
value, which corresponds to the fractional D-branes moving around 
independently on the ALE singularity. 
The bi--fundamental hyper's parameterise the Higgs branch, describing the
motions of the D1--brane away from the ALE singularity. 
Naively there ought to be
$Q_m^2$ strings corresponding to the number of ways of connecting
together the D--branes, but in fact we only have $Q_m$ massless
sectors. The latter
arises as the solution to the constraints on the Chan--Paton
factors placed by the $\mathbb{Z}_{Q_m}$ symmetry. A similar result occurs
for the more complicated case of D--branes made of D1--D5 bound states
which we will show gives the simple result for the entropy.

In order to generalize these results to the present problem we begin
by considering the effective 1+1 dimensional theory living on a
single D1--D5 bound state\footnote{This is a misnomer for just a
  single D1--brane and D5--brane, but the rest of the discussion
  carries through. When there are more D5--branes, the term ``bound
  state'' will be accurate, since then we can form instantons, as we
  have a non--Abelian gauge group on their world--volume.}  sitting
at the origin of the $\mathbb{Z}_{Q_m}$ orbifold with $Q_1=Q_5=1$.  The
spectra of $(1,1)$ and $(5,5)$ strings are given by two copies of the
spectra found in the single D1--brane case. In particular the gauge
group is $[U_1(1)\times U_5(1)]^{Q_m}$ where the subscripts $1,5$
refer to which brane the gauge group originated on. Here the $(1,1)$
strings give a gauge field and adjoint hyper of $U_1(1)^{Q_m}$ and a
set of hyper-multiplets in the bi--fundamentals of the adjacent factors
of $U_1(1)$.  These are all singlets of the $U_5(1)$ factors. In
exactly the same way there is a vector, adjoint hyper-multiplets and a
set of bi--fundamental hyper-multiplets coming from the $(5,5)$
strings.  So far therefore, we have simply made two copies of the
non--standard (or quiver) gauge theory arising from D--branes in the presence
of an ALE singularity.  The Higgs branch is given by the vev's of the
bi--fundamentals,
making the branes move off the ALE space, and the Coulomb by
the adjoint scalars, allowing the fractional D1-branes to move around inside
it. In the $U_5(1)^{Q_m}$ theory, the adjoint scalar vev's can be thought of as
introducing Wilson
lines on the $T^4$ in a six dimensional gauge theory on the wrapped D5-branes.

The novelty here is that there is another sector, because there are
$(1,5)$ and $(5,1)$ strings. Naively there are quite a few of these,
since there are many ways of connecting each of the $Q_m$ D1--branes
to the $Q_m$ D5--branes. There are far fewer in fact, and to see
this we will consider them in some detail. 

In order to facilitate the analysis we recall that the bound state we
are considering here breaks the space-time Lorentz symmetry as,
$SO(1,9)\rightarrow SO(1,1)\times SO(4)_E\times SO(4)_I$ where the
$SO(1,1)$ factor represents Lorentz rotations in the $(09)$
directions, the $SO(4)_E$ factor is the rotations in the non--compact
external directions $(1234)$ and the factor $SO(4)_I$ are the
rotations of the internal compact directions $(5678)$.  The worldsheet
fields on these strings have mixed boundary conditions. Specifically,
oscillators in the (09) directions have Neumann--Neumann (NN) boundary
conditions, oscillators in the (1234) directions have
Dirichlet--Dirichlet (DD) boundary conditions while the (5678)
oscillators have ND/DN boundary conditions. This system has four ND
boundary conditions and thus the zero point energy of oscillators in
both the R and NS sectors vanish (this is reviewed in
refs.~\cite{joe2,cvj}). In the ND directions the the NS sector states are
periodic and thus have zero modes $\psi_0^i, i=5,6,7,8$ while the NN/DD
directions have a periodic R sector with zero modes $ \psi_0^m, 
m=0,1,2,3,4,9$. 
After the GSO projection we are left, from the 1+1 dimensional
point of view, with a single Weyl spinor and
two real scalars, or half of a hyper-multiplet.  Noting that there is a
doubling of states due to the reversed orientation of $(1,5)$ and
$(5,1)$ strings we see that there is an entire hyper-multiplet.
For our purposes below we remind the reader that the Weyl spinors 
of this hyper-multiplet can be written in
the Chevalier basis as  $\chi=|s_1s_2>$ where  
$s_1,s_2$ are the eigenvalues of the Cartan generators 
of $SO(4)_E$, which we will denote as $S_1,S_2$ below.
Further imposing the GSO projection selects $s_1=s_2$.

In order to determine the spectrum of these strings when this system
is placed on the $\mathbb{Z}_{Q_m}$ orbifold we need to consider how the
orbifold acts on these oscillators.  Clearly the
ND (\ie $\psi_{0}^{i}$) ground states are invariant.
For the $\psi_{0}^{m}$ ground states, we
adopt the conventions of ref.~\cite{eric1}.  Specifically, we assemble
the relevant worldsheet fields into complex pairs. We define
$z_1=X^1+iX^2$ and $z_2=X^3+iX^4$ for the bosonic fields and likewise
for the fermions. Denoting the elements of $\mathbb{Z}_{Q_m}$ as
$\alpha_{Q_m}^k$ where $k=1,2\cdots Q_m$ the action on the bosonic
and NS sector fields is then given as,
\begin{equation}
z_1\rightarrow \alpha_{Q_m}^k z_1= e^{\frac{2\pi ik}{Q_m}}z_1 \,\,\,\,\,\, 
z_2\rightarrow \alpha_{Q_m}^k z_2= e^{-\frac{2\pi ik}{Q_m}}z_2
\labell{act}
\end{equation}
while for the R sector fermions we have\footnote{To motivate this choice of 
action on R sector states see refs.~\cite{eric2,eric1}.}
\begin{equation}
\alpha_{Q_m}^k=e^{\frac{2\pi ik}{Q_m}(S_1-S_2)}
\labell{Ract}
\end{equation}
Thus, due to the GSO projection, the action of
$\mathbb{Z}_{Q_m}$ on the R sector ground state is trivial. 

We can now determine the massless spectrum of $(1,5)$ strings.
Following ref.~\cite{ale} we note that when the D1--D5 system is
brought to the fixed point of the orbifold so too are all of its
images and hence the naive gauge group is $U_1(Q_m)\times
U_5(Q_m)$. As a result the open strings now carry Chan--Paton factors
which we represent by $Q_m\times Q_m$ matrices
$\lambda^m_{ij}$.
As in ref.~\cite{ale} the Chan Paton
indices must be invariant under the orbifold up to its action on the space
time indices, which as we have just shown above is trivial for both
the ND and DD strings. The constraint that we require is
written as,
\begin{equation}
\tilde{\lambda}_{ij}=\gamma_{im}\lambda_{mn}(\gamma)^{-1}_{nj}=\lambda_{ij}
\labell{inva}
\end{equation}
where $\gamma_{ij}$ is the $Q_m$ dimensional regular 
representation\footnote{Recall that we can always choose a basis in which the
regular representation is block diagonal where each irreducible
representation of dimension $n$ appears $n$ times along the diagonal
~\cite{group}. Here the regular representation is simply given by a
diagonal matrices whose entries are the $Q_m$ roots of unity.} of 
$\mathbb{Z}_{Q_m}$.
This can only be satisfied by diagonal matrices. In particular there are 
{\it no} massless $(1,5)$ strings connecting different factors of the
gauge group, {\it i.e., there are no $(1,5)$ strings linking different
  images of the D1--D5 system}. There are then a total of $Q_m$
hyper-multiplets, one for each {\it non-trivial} factor of
$U(1)_1\times U_5(1)$. Since the GSO projection left us with two Weyl
spinors of $SO(4)_E$ (one for $(1,5)$ and one for $(5,1)$) we see that
there are $4Q_m$ boson/fermion ground states corresponding to these
hyper-multiplets.

We now generalize this discussion to arbitrary $Q_1$ and $Q_5$.  To do
so we first discuss the fate of the $(1,1)$ (or equivalently the
$(5,5)$) strings in order to determine what the gauge group is.  For
the case where $Q_1$ D-strings are brought to the fixed point of an
$A_{Q_m-1}$ orbifold the gauge group is $U(Q_mQ_1)$ before
projecting onto invariant states. To determine the surviving gauge
group we note that the Chan-Paton factors of the gauge fields,
$A_{\mu}$ with $\mu =0,9$ are now $Q_mQ_1\times Q_mQ_1$
hermitian matrices $\Lambda$. The constraint imposed by the orbifold on the
Chan--Paton matrices is,
\begin{equation}
\tilde{\Lambda}_{ab}=(\gamma_{Q_1})_{ac}\Lambda_{cd}(\gamma_{Q_1})^{-1}_{db}=
\Lambda_{ab}
\labell{inv3}
\end{equation}
where $\gamma_{Q_1}$ is the $Q_mQ_1$ dimensional regular
representation of $\mathbb{Z}_{Q_m}$ formed by tensoring the $Q_{m}$
dimensional representation used in eqn.~\reef{inva} with the $Q_1\times
Q_1$ identity matrix.  This condition can only be solved by block
diagonal $\Lambda$ whose blocks are $Q_1\times Q_1$ dimensional
hermitian matrices. The unbroken gauge group is thus $U(Q_1)^{Q_m}$.
The same reasoning applies to the $(5,5)$ strings in the D1--D5 system
except in this case one constructs the regular representation of
$\mathbb{Z}_{Q_m}$ using the $Q_5\times Q_5$ identity matrix.  The unbroken
gauge group when a D1--D5 system comprised of $Q_1$ D1--branes and $Q_5$
D5--branes is put on top of an $A_{Q_m-1}$ singularity is thus
$[U(Q_1)\times U(Q_5)]^{Q_m}$.  The matter content from the $(1,1)$
and $(5,5)$ strings can be determined exactly as outlined for the
gauge bosons above: There are a pair of adjoint hyper-multiplets one of
which is in the adjoint of $U(Q_1)^{Q_m}$ and the other is in the
adjoint of $U(Q_5)^{Q_m}$; there will also be $Q_m$
hyper-multiplets each transforming in the fundamental representation of
the $U(Q_1)$ factors which correspond to fundamental strings 
stretched between
D1--branes in different sectors, likewise there will be $Q_m$
hyper-multiplets transforming in the fundamentals of the $U(Q_5)$
factors which represent strings stretched between D5--branes in
different sectors.

Finally we come to the $(1,5)$ strings. The answer is found by
combining the analyses of the previous paragraphs. Before performing the
orbifold projection these strings are in the bi--fundamental
representation of $U(Q_mQ_1)\times U(Q_mQ_5)$. As shown before
the oscillators for these states are acted on trivially by $\mathbb{Z}_{Q_m}$
thus the constraint is simply a generalization of that presented in
eqn.~\reef{inv3}.  The Chan--Paton factors of these states are now
$Q_mQ_1\times Q_mQ_5$ matrices $\Omega$ which can be thought
of as $Q_m\times Q_m$ matrices whose entries are $Q_1 \times
Q_5$ blocks.  Following eqn.~\reef{inv3} we impose the constraint,
\begin{equation}
\tilde{\Omega}_{aB}=(\gamma_{Q_1})_{ac}\Omega_{cD}(\gamma_{Q_5})^{-1}_{DB}=
\Omega_{aB}
\labell{invar4}
\end{equation}
where $\gamma_{Q_1}$ is as before, $\gamma_{Q_5}$ is formed by tensoring
the $Q_m$ dimensional regular representation of $\mathbb{Z}_{Q_m}$ with the
$Q_5\times Q_5$ identity matrix. Here lower case indices run over
$1,\cdots, Q_1$ while upper case indices run over $1,\cdots, Q_5$. This
constraint can only be solved by ``diagonal'' matrices whose entries are
$Q_1 \times Q_5$ blocks. There are therefore {\it no} strings in
bi--fundamental representations of different factors of the gauge group.
More succinctly there are no $(1,5)$ strings in the links of the
Dynkin diagram. Thus the only $(1,5)$ strings are those which connect
D1--branes and D5--branes in the same sector of the gauge theory. So our
conclusion is that there are precisely $Q_m$ hyper-multiplets, one
for each factor appearing in the gauge group, corresponding to $(1,5)$
and $(5,1)$ strings. See figure~\ref{fig:quiv}b.

Note that the low energy effective 
field theory discussed here will flow
  to a 1+1 dimensional conformal field
theory in the IR which is dual to the near horizon geometry \reef{near}
discussed in section (2) --- see refs.~\cite{bern,suga}.
We also refer to the reader to ref.~\cite{leigh2} for a discussion
of the closely related gauge theory appearing in the bound state
of two D3--branes on an ALE singularity.

Now to give a microscopic account of the black hole entropy, we calculate
the ground state degeneracy when $N$ units of momentum are introduced
in the 1+1 dimensional gauge theory. There are 
a total of $Q_m$ hyper-multiplets coming from $(1,5)$ strings, each
of which has $4Q_1Q_5$ boson/fermion ground states making a total of
$4Q_mQ_1Q_5$.  The degeneracy of states then comes from the number of
ways we can distribute $N$ units of momentum to these degrees of
freedom. Since this system is BPS, we can easily write the partition
function for the momentum $N$ among these bosonic and fermionic states
in the 1+1 dimensional effective theory as a purely chiral system:
\begin{equation}
\prod_{N=1}^\infty\left( {1+q^N\over 1-q^N
    }\right)^{4Q_mQ_1Q_5}=\sum_{N=1}^{\infty} \Omega(N) q^N\ ,
\labell{answer3}
\end{equation}
where, at large $N$, the level degeneracy behaves as $\Omega(N)\sim
\exp(2\pi\sqrt{Q_mQ_1Q_5N}).$ We see that the entropy is therefore
\begin{equation}
S=\log\Omega(N)=2\pi\sqrt{Q_mQ_1Q_5N}
\labell{answer2}
\end{equation}
which is precisely that given in eqn.~\reef{entropy}!

\section{Discussion}
So we have our desired result: The entropy of the four dimensional
extremal black hole constructed from D1--branes, D5--branes, momentum
and a Kaluza--Klein monopole, is indeed the product of four charges
(associated to each sector) appropriate for the quartic invariant of
$E_{(7)7}$. This has of course been shown elsewhere~\cite{sm} using
a different construction in the type IIA theory,
not involving Kaluza--Klein monopoles, but
related to the one presented here by U--duality.  It is however, quite
satisfying to show it explicitly with the original arrangement of
constituents, and pleasing that the interpretation in terms of
fractional branes involving the machinery of D--branes on ALE spaces
manifests itself so clearly. In particular, the absence of $(1,5)$
strings stretching between different factors of the gauge group is
quite dramatic, since had it been the naive result, the entropy would
have acquired a factor of $Q_m$, instead of the required $Q_m^{1/2}$,
which is a significant difference at large $Q_m$.
  
Note that the expression \reef{answer2} cannot be the full 
U--duality invariant expression for the counting of the BPS states
in this system.
In particular, one might note that our result is not valid
for small $N$. In general, we might expect
corrections to appear for small values of the charges both in the
gauge theory calculations and in the black hole entropy. The
latter can arise from higher curvature corrections to the supergravity
action, which will modify the form of the black hole solution
(and hence its horizon area), and also introduce corrections to the
Bekenstein-Hawking entropy
formula \cite{wald} --- for a general review, see ref.~\cite{wald2}.
In certain cases, such subleading corrections to the entropy have
been matched in microscopic string calculations for black holes with 
${\cal N} =2$ supersymmetry \cite{msw,vafa2,dewitt}.
Generically, one also finds logarithmic corrections \cite{log2} to black hole
entropy in going beyond the large charge approximation used, \eg in deriving
eqn.~\reef{answer2} from the previous expression \reef{answer3}.
It would be interesting to extend our calculations
to determine the full U--duality invariant result using techniques
analogous to those used in refs.~\cite{farey}.\footnote{We thank
J. Maldacena for correspondence on this point.}

Given the success
of those techniques above in examining the D1--D5 bound state, it seems that
the methods used here are likely to be useful in extending these
results to a U--dual counting formula of the BPS states underlying 
the present four
dimensional black holes. However, one should note that the most general
supersymmetric configuration in four dimensions actually depends on {\it five}
charges\cite{gener1,berta}. That is while four charges is the minimum
required to produce a non-vanishing value for the quartic invariant of
$E_{(7)7}$, the generating solution from which one can obtain all
other configurations by the action of U--duality actually carries five
independent charges. Recently progress has been made in producing a
microscopic description of the black hole entropy of such five charge
configurations \cite{bert} by applying the approach of ref.~\cite{msw}
to an M-theory embedding of the solution. The simplest approach
to adding a fifth charge to the present
type IIB black holes would require introducing an additional momentum
in the $x^4$ direction, as well as a diagonal wrapping of either
the D1-- or D5--branes on this direction. Certainly extending
the present analysis to this more elaborate configuration would be
a necessary first step in producing a complete U--duality invariant
expression for the black hole entropy.

\section{Acknowledgments}
Research by RCM and NRC was supported by NSERC of Canada and Fonds FCAR du
Qu\'ebec. RCM and CVJ would like to thank the Aspen Center for Physics
for hospitality during part of this work.
We would also like to thank \O yvind Tafjord for useful conversations.

\end{document}